\documentclass[amssymb,amsmath,aps,showpacs,floatfix,nofootinbib,showpacs,12pt]{revtex4}
\usepackage{epsfig}
\usepackage{color}
\usepackage{graphicx}

\def\beq{\begin{equation}}
\def\eeq{\end{equation}}

\def\be{\begin{equation}}
\def\ee{\end{equation}}
\def\bea{\begin{eqnarray}}
\def\eea{\end{eqnarray}}

\newcommand{\gsim}{\lower.7ex\hbox{$\;\stackrel{\textstyle>}{\sim}\;$}}
\newcommand{\lsim}{\lower.7ex\hbox{$\;\stackrel{\textstyle<}{\sim}\;$}}

\begin{document}

\hspace{4.5in}IPMU 10-0060

\bigskip

\title{Hunting for New Physics with Unitarity Boomerangs
}

\author{Paul H. Frampton$^{1,2}$ and Xiao-Gang He$^{3,4}$}
\affiliation{
$^1$Department of Physics and Astronomy,
University of North Carolina, Chapel Hill, NC 27599-3255, USA\\
$^2$Institute for the Physics and Mathematics of the Universe,
University of Tokyo, Kashiwa, Chiba 277-8568, JAPAN\\
$^3$Department of Physics and Center for Theoretical Sciences, National Taiwan University, Taipei\\
$^4$Institute of Particle Physics and Cosmology, Department of Physics,
Shanghai JiaoTong University, Shanghai}

\begin{abstract}
Although the unitarity triangles ($UTs$) carry information about the Kobayashi-Maskawa (KM) quark mixing matrix, it explicitly
contains just three parameters which is one short to completely fix the KM matrix. It has been shown recently, by us, that the unitarity boomerangs ($UB$)
formed using two $UTs$, with a common inner angle, can completely determine the KM matrix and, therefore, better represents, quark mixing. Here, we study detailed properties of
the $UBs$, of which there are a total 18 possible. Among them, there is only one which does not involve very small angles and is the ideal one for practical uses.
Although the $UBs$ have different areas, there is an invariant quantity, for all $UBs$, which is equal to
a quarter of the Jarlskog parameter $J$ squared. Hunting new physics, with a unitarity boomerang, can reveal more information, than just using a unitarity triangle.
\end{abstract}


\maketitle

\noindent
{\bf Introduction}

The Kobayashi-Maskawa~\cite{KM} (KM) quark mixing matrix, $V_{KM}$, can describe all laboratory data on quark mixing and CP violation.
There are different ways of parameterizing the KM matrix. For three generations of quarks,
$V_{KM}$ is a $3\times 3$ unitary mixing matrix with three rotation angles $(\theta_1,\;\theta_2,\;\theta_3)$ and one CP violating
phase $\delta$. The magnitudes of the elements $(V_{KM})_{ij}$ of $V_{KM}$
are physical quantities which do not depend on parametrization. However, the value of $\delta$ does~\cite{PDG,ck,xing2,fh}. Care must be exercised in quoting a value of $\delta$,
as it depends on how the matrix is parameterized.
It is therefore desirable to employ only physically-measurable quantities. To this end, it has long ago been suggested that
a unitarity triangle (UT) be used\cite{BJ} as a useful presentation for
the quark flavor mixing, especially of CP violation~\cite{CCFT}.

\bigskip

Because the unitary nature of the KM matrix, the elements $(V_{KM})_{ij}$ in the matrix satisfy
\begin{eqnarray}
\sum_i (V_{KM})_{ij}(V_{KM})^*_{ik} = \delta_{jk}\;,\;\;\sum_i (V_{KM})_{ji}(V_{KM})^*_{ki} = \delta_{jk},
\end{eqnarray}
where the first and second indices of $(V_{KM})_{ij}$
take the values $u,\;c,\;t,\;...$ and $d,\;s,\;b,\;...$, respectively.
For three generations of quarks, when $j\neq k$, these equations define six triangles in a plane, the $UTs$.
All of the six $UTs$ have the same area. $A(UT)$, which is equal to half of the value of the Jarlskog
parameter~\cite{jarlskog} $J$, so that $A(UT)=J/2$.
The inner angles of a given $UT$ are therefore closely related
to the CP violating measure $J$. When the inner angles are measured independently,
their sum, whether it turns out to be consistent with precisely
$180^\circ$, provides a test for the unitarity of the KM matrix. The unitarity triangle is a popular
way, to present CP violation, with three generations of quarks.

\bigskip

A $UT$, however, does not contain all the information encoded in the KM matrix, $V_{KM}$.
Although a $UT$ has three inner angles and three sides, it contains only three independent parameters. The three parameters
can be chosen to be two of the three inner angles and the area, or the three sides, or some combination thereof.
One needs an additional parameter fully to represent the physics contained in the KM matrix. This is not a surprise because the
$UT$ involved only two, of the three, rows or columns of the $3 \times 3$ matrix, $V_{KM}$,

\bigskip

An improved presentation is thus rendered desirable, in order better
to present the KM matrix, $V_{KM}$. We have, in a recent paper~\cite{fh}, proposed
a new graphic representation of the KM matrix by using the boomerang diagram, the unitarity boomerang ($UB$).
The boomerang diagram  contains information from not just one $UT$, but two $UTs$, from among the six possible different $UTs$.
Some extensions of $UT$ analysis to lepton section
has been considered in the Ref.~\cite{lm}.

\bigskip

Out of the six possible $UTs$, there are 9 different ways to have a common inner angle from two $UTs$. In addition one can
align the two longer sides or one longer and one shorter sides to form a $UB$. Therefore total 18 possible $UBs$.
We find that among these $UBs$, there is only one which does not involve very small angles and is good for practical uses.
Although the $UBs$ have different areas, there is an invariant quantity for all $UBs$ which is equal to
a quarter of the Jarlskog parameter $J$ squared. We also discuss how new physics can be tested by using information from $UBs$.

\bigskip

\noindent
{\bf The Unitarity Triangles and Boomerangs}

The $UBs$ are constructed by using two $UTs$.  Let us summarize some of the relevant information for $UTs$ here.
There are six triangles. The inner angles of three $UTs$ involving two columns are defined by, $\sum_i (V_{KM})_{ij}(V_{KM})^*_{ik} = 0\;, \mbox{$j\neq k$}$,
\begin{eqnarray}
UT(1):\;\;&&(V_{KM})_{ub}(V_{KM})^*_{us} + (V_{KM})_{cb}(V_{KM})^*_{cs} + (V_{KM})_{tb}(V_{KM})^*_{ts} = 0\;,\nonumber\\
&&\alpha_1 = arg\left (-{(V_{KM})_{ub}(V_{KM})_{us}^*\over (V_{KM})_{cb}(V_{KM})^*_{cs}}\right )\;,\;\beta_1 = arg \left (-{(V_{KM})_{tb}(V_{KM})^*_{ts}\over (V_{KM})_{ub}(V_{KM})^*_{us}}\right )\;,
\nonumber\\
&&\gamma_1 = arg \left (-{(V_{KM})_{cb}(V_{KM})^*_{cs}\over (V_{KM})_{tb}(V_{KM})_{ts}^*}\right )\;,\nonumber\\
&&\nonumber\\
UT(2):\;\;&&(V_{KM})_{ud}(V_{KM})^*_{ub} + (V_{KM})_{cd}(V_{KM})^*_{cb} + (V_{KM})_{td}(V_{KM})^*_{tb} = 0\;,\nonumber\\\
&&\alpha_2 = arg \left (-{(V_{KM})_{td}(V_{KM})_{tb}^*\over (V_{KM})_{ud}(V_{KM})^*_{ub}}\right )\;,\;\beta_2 = arg \left (-{(V_{KM})_{cd}(V_{KM})^*_{cb}\over (V_{KM})_{td}(V_{KM})^*_{tb}}\right )\;,
\nonumber\\
&&\gamma_2 = arg \left (-{(V_{KM})_{ud}(V_{KM})^*_{ub}\over (V_{KM})_{cd}(V_{KM})_{cb}^*}\right )\;,\nonumber\\
&&\nonumber\\
UT(3):\;\;&&(V_{KM})_{ud}(V_{KM})^*_{us} + (V_{KM})_{cd}(V_{KM})^*_{cs} + (V_{KM})_{td}(V_{KM})^*_{ts} = 0\;,\nonumber\\
&&\alpha_3 = arg \left (-{(V_{KM})_{td}(V_{KM})^*_{ts}\over (V_{KM})_{cd}(V_{KM})^*_{cs}}\right )\;,
\;\;\beta_3 = arg \left (-{(V_{KM})_{cd}(V_{KM})^*_{cs}\over (V_{KM})_{ud}(V_{KM})_{us}^*}\right )\;,\nonumber\\
&&\gamma_3 = arg \left (-{(V_{KM})_{ud}(V_{KM})_{us}^*\over (V_{KM})_{td}(V_{KM})^*_{ts}}\right )\;.
\label{ut1}
\end{eqnarray}
For convenience, we have labeled the $UT(i)$ by the missing $i$th quark in the down-quark sector.

\bigskip

The inner angles of the three UTs involving two rows are given by, $\sum_i (V_{KM})_{ji}(V_{KM})^*_{ki} = 0\;, \mbox{$j\neq k$}$,
\begin{eqnarray}
UT(1'):\;\;&&(V_{KM})_{cd}(V_{KM})^*_{td} + (V_{KM})_{cs}(V_{KM})^*_{ts} + (V_{KM})_{cb}(V_{KM})^*_{tb} = 0\;,\nonumber\\
&&\alpha'_1 = arg \left (-{(V_{KM})_{cs}(V_{KM})^*_{ts}\over (V_{KM})_{cd}(V_{KM})^*_{td}}\right )\;,\;\;\beta'_1 = arg \left (-{(V_{KM})_{cd}(V_{KM})_{td}^*\over (V_{KM})_{cb}(V_{KM})^*_{tb}}\right )\;,\nonumber\\
&&\gamma'_1 = arg \left (-{(V_{KM})_{cb}(V_{KM})^*_{tb}\over (V_{KM})_{cs}(V_{KM})_{ts}^*}\right )\;,\nonumber\\
&&\nonumber\\
UT(2'):\;\;&&(V_{KM})_{ud}(V_{KM})^*_{td} + (V_{KM})_{us}(V_{KM})^*_{ts} + (V_{KM})_{ub}(V_{KM})^*_{tb} = 0\;,\nonumber\\
&&\alpha'_2 = arg \left (-{(V_{KM})_{ub}(V_{KM})_{tb}^*\over (V_{KM})_{ud}(V_{KM})^*_{td}}\right )\;,\beta'_2 = arg \left (-{(V_{KM})_{us}(V_{KM})^*_{ts}\over (V_{KM})_{ub}(V_{KM})^*_{tb}}\right )\;,\nonumber\\
&&\gamma'_2 = arg \left (-{(V_{KM})_{ud}(V_{KM})^*_{td}\over (V_{KM})_{us}(V_{KM})_{ts}^*}\right )\;,\nonumber\\
&&\nonumber\\
UT(3'):\;\;&&(V_{KM})_{ud}(V_{KM})^*_{cd} + (V_{KM})_{us}(V_{KM})^*_{cs} + (V_{KM})_{ub}(V_{KM})^*_{cb} = 0\;,\nonumber\\
&&\alpha'_3 = arg \left (-{(V_{KM})_{ub}(V_{KM})^*_{cb}\over (V_{KM})_{us}(V_{KM})^*_{cs}}\right )\;,\;\;\beta'_3 = arg \left (-{(V_{KM})_{us}(V_{KM})^*_{cs}\over (V_{KM})_{ud}(V_{KM})_{cd}^*}\right )\;,\nonumber\\
&&\gamma'_3 = arg \left (-{(V_{KM})_{ud}(V_{KM})_{cd}^*\over (V_{KM})_{ub}(V_{KM})^*_{cb}}\right )\;.
\label{utp}
\end{eqnarray}
Here the $UT(i')$ labeled by the missing $i'$th quark in the up-quark sector.

\bigskip

It is clear from the above definitions that among the 18 inner angles of the six $UTs$,
only 9 of them are different. Explicitly, we have
\begin{eqnarray}
&&\alpha_1 = \alpha'_3\;,\;\;\beta_1 = \beta'_2\;,\;\;\gamma_1 = \gamma'_1\;,\nonumber\\
&&\alpha_2 = \alpha'_2\;,\;\;\beta_2 = \beta'_1\;,\;\;\gamma_2 = \gamma'_3\;,\nonumber\\
&&\alpha_3 = \alpha'_1\;,\;\;\beta_3 = \beta'_3\;,\;\;\gamma_3 = \gamma'_2\;.
\label{boomer}
\end{eqnarray}
We will choose the 9 inner angles without ``prime'' in our later discussions.

\bigskip

As pointed out earlier that for a given $UT$, it contains only 3 independent parameters which is not enough to completely determine parameters in the KM matrix.
In a previous paper~\cite{fh}  we have shown that using two triangles with one from $UT(i)$ and another from $UT(i')$ one can always form a boomerang like diagram, the unitarity boomerang and ($UB$) contains all information need to reconstruct the KM matrix. Let us show more details in the following.

\bigskip

There are total 18 different ways of constructing $UBs$. There are 9 different ways to pair up a common angle by taking one $UT$ from $UT(i)$ and another from $UT(i')$. One can then overlap the longer side from on $UT$ with the shorter side of the other $UT$, or overlap the longer sides of the two $UTs$.  We will label the former 9 $UBs$ as $B_{ii'a}$ and the later 9 $UBs$ as $\tilde B_{ii'a}$. Here the index $a$ indicates the common angle.
The common angles and their current central experimental values of the $UBs$ are given in the following
\begin{eqnarray}
&&\left ( \begin{array}{lll}
B(\tilde B)_{11\gamma_1}&B(\tilde B)_{12\beta_1}&B(\tilde B)_{13\alpha_1}\\
B(\tilde B)_{21\beta_2}&B(\tilde B)_{22\alpha_2}&B(\tilde B)_{23\gamma_2}\\
B(\tilde B)_{31\alpha_3}&B(\tilde B)_{32\gamma_3}&B(\tilde B)_{33\beta_3}
\end{array} \right)\nonumber\\
\Rightarrow &&
\left (\begin{array}{lll}
\gamma_1 = 0.0207&\beta_1=1.151&\alpha_1 = 1.970\\
\beta_2= 0.435 & \alpha_2 = 1.535 &\gamma_2 = 1.171 \\
\alpha_3 = 2.686&\gamma_3 = 0.455 &\beta_3 = 5.531\times 10^{-4}
\end{array} \right )\;.
\end{eqnarray}
In the above when calculating the inner angles, we have used the central values~\cite{PDG} $|(V_{KM})_{ud}| = 0.97419$, $|(V_{KM})_{us}| = 0.2257$, $|(V_{KM})_{ub}| = 0.00359$ and $\alpha_2 = 88^\circ$. We will use these values in our later discussions for illustrations.

\bigskip

We now display the $UBs$. In Fig. 1 we show some details for the boomerang formed by using $UT(2)$ and $UT(2')$. Figs. 1. a) and 1. b) are for $B_{22\alpha_2}$, and $\tilde B_{22\alpha_2}$.  The 9 $B_{ii'a}$ are shown in Fig. 2. One can also construct the 9 $UBs$ of $\tilde B_{ii'a}$. But they contain similar information as those of $B_{ii'a}$, they can be obtained by flipping one of the $UT$ in each of the $UB$, we therefore have not displayed them.

\begin{figure}[hb]
\includegraphics[width=6in]{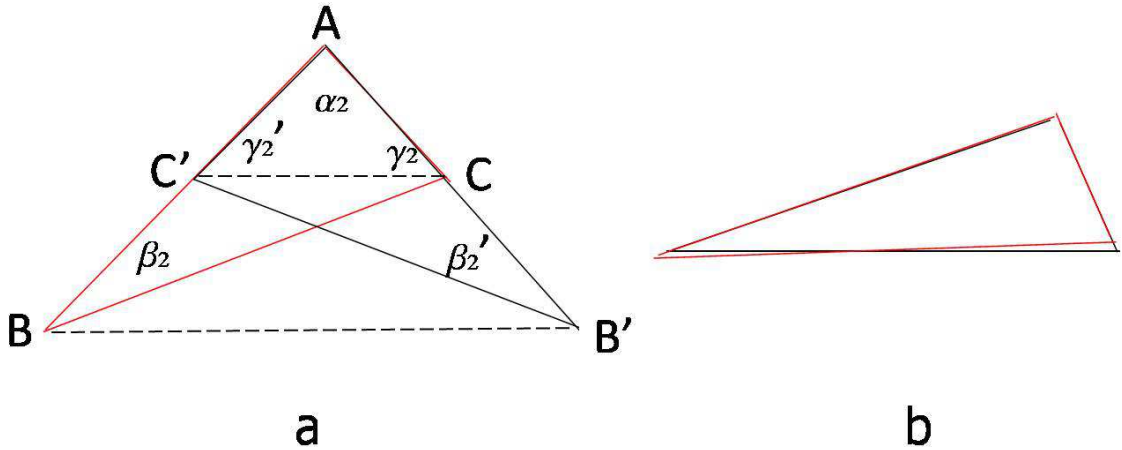}
\caption{The unitarity boomerangs $B_{22\alpha_2}$ and $\tilde B_{22\alpha_2}$. Figure 1.a) overlaps one long and one short sides from the two $UTs$. The sides are: $AB = |(V_{KM})_{td}(V_{KM})^*_{tb}|$, $AC = |(V_{KM})_{ud}(V_{KM})^*_{ub}|$, $BC = |(V_{KM})_{cd}(V_{KM})^*_{cb}|$, $AB' = |(V_{KM})_{ud}(V_{KM})^*_{td}|$,  $AC' = |(V_{KM})_{ub}(V_{KM})^*_{tb}|$, and $B'C' = |(V_{KM})_{us}(V_{KM})^*_{ts}|$. Figure 1.b) is for $\tilde B_{22\alpha}$. The red (left) lines and black (right) lines indicate the $UT$ taking from $UT(i)$ and $UT(i')$, respectively.
\label{boomerang}}
\end{figure}

\begin{figure}[hb]
\includegraphics[width=5in]{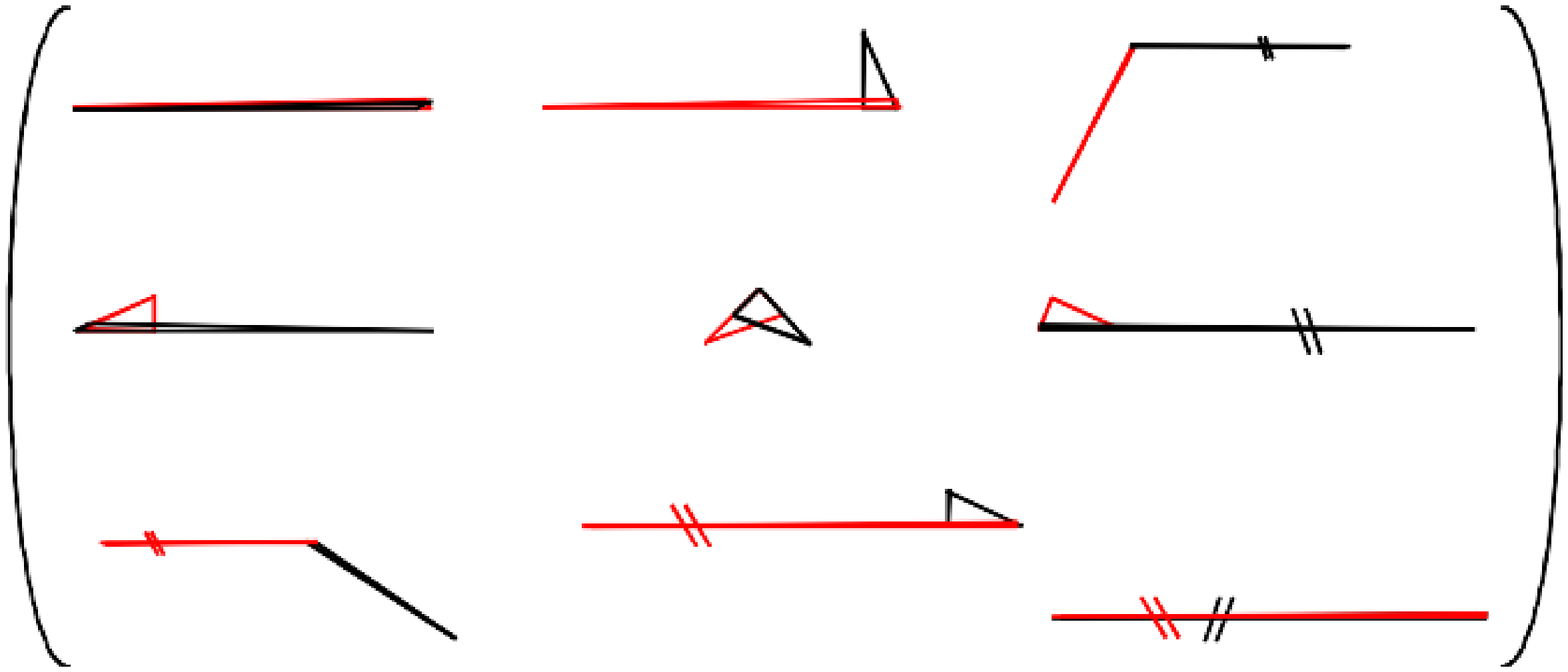}
\caption{The 9 unitarity boomerangs $B_{ii'a}$.  The red (left) lines and black (right) lines indicate the $UTs$ taking from $UT(i)$ and $UT(i')$. In the figure, the symbols double back-slash and double slash indicate the $UT$ for $i=3$ and $i'=3$.
For the 23, 32, 33 entries, the length in the figure for $UT(3)$ and $UT(3')$ should be scaled up by 5. For the 13 and 31 entries, the triangles involve $UT(3,3')$ and $UT(1,1')$ should  be scaled up by 10 and 2, respectively.
\label{boomerang}}
\end{figure}

As can be seen from Fig. 2, the $UBs$, except the $B_{22\alpha_2}$, all involve a very small angle in the diagram making it difficult to construct, with high accuracy. The $B_{22\alpha_2}$
can display the information more easily. One should therefore work with the $B_{22\alpha_2}$. From the $B_{22\alpha_2}$, one can easily obtain approximate solutions for the
four physical parameters.
Taking the ratio, of the two sides $AC/AC'$ or $AB/AB'$, one obtains, $|(V_{KM})_{ud}/(V_{KM})_{tb}^*| \approx c_1$
since $|(V_{KM})_{tb}|$ is very close to 1. With $c_1$ and therefore $s_1$ known, the length of the sides AB and AC' then provide the values for $s_2$ and $s_3$. Here the angles $c_i$ and $s_i$ refer to the cosine and sine of the angles in the original KM matrix parametrization. One can also obtain more accurate expressions, as shown in Ref.\cite{fh}. In this parametrization, the CP violating phase $\delta$ is, to a very good approximation, equal to~\cite{koide2,fh} $\alpha_2$.

\bigskip

\noindent
{\bf Invariant quantity for CP violation}

Information on CP violation are, also, fully encoded, in the $UBs$.
The Jarlskog parameter $J$ plays a fundamental role for CP violation with three generations. In the $UT$ representation, it is related to the area of the $UT$. All $UTs$ have the same
area of $J/2$. One can also construct a geometric representation of CP violation invariant quantity for $UB$ representation.  Although the shapes of the $UBs$ vary a lot, there exists  a quantity related to the $UB$ constructed areas
represent CP violation in an invariant form.

\bigskip

Consider the areas $A_{ABB'}$ and $A_{ACC'}$ of the two triangles $ACC'$ and $ABB'$ in Fig. 1 for $B_{22\alpha_2}$, we find
\begin{eqnarray}
A_{22} = A_{ABB'} = \frac{1}{2}\frac{|(V_{KM})_{td}|}{|(V_{KM})_{ub}|}J\;,\;\;A'_{22} = A_{ACC'} = \frac{1}{2}\frac{|(V_{KM})_{ub}|}{|(V_{KM})_{td}|}J\;;\;\;A_{22}A'_{22} = \frac{1}{4}J^2.
\end{eqnarray}
The two triangles $ABB'$ and $ACC'$ are similar to each other.

\bigskip

Similarly for $\tilde B_{22\alpha_2}$, one has
\begin{eqnarray}
\tilde A_{22} = \tilde A_{\tilde A \tilde B \tilde B'} = \frac{1}{2}\frac{|(V_{KM})_{td}|}{|(V_{KM})_{ub}|}J\;,\;\;\tilde A'_{22} = \tilde A_{\tilde A \tilde C \tilde C'} = \frac{1}{2}\frac{|(V_{KM})_{ub}|}{|(V_{KM})_{td}|}J\;,\;\;\tilde A_{22} \tilde A'_{22} = \frac{1}{4}J^2\;.
\end{eqnarray}
Again the two triangles are similar to each other.

\bigskip

This can be generalized to all  $UBs$. The results can be collected in matrix forms. Writing $A_{22}$, $A'_{22}$, $\tilde A_{22}$ and $\tilde A'_{22}$ like
areas for other $UBs$ in similar ways, the resultant matrix forms $A$, $A'$, $\tilde A$ and $\tilde A'$ are given by
\begin{eqnarray}
&&A = \frac{1}{2}  \left ( \begin{array}{lll}
\frac{|(V_{KM})_{tb}|}{|(V_{KM})_{cs}|}&\frac{|(V_{KM})_{ts}|}{|(V_{KM})_{ub}|}&\frac{(V_{KM})_{cs}|}{|(V_{KM})_{ub}|}\\
\frac{|(V_{KM})_{cb}|}{|(V_{KM})_{td}|}&\frac{|(V_{KM})_{td}|}{|(V_{KM})_{ub}|}&\frac{(V_{KM})_{cd}|}{|(V_{KM})_{ub}|}\\
\frac{|(V_{KM})_{cs}|}{|(V_{KM})_{td}|}&\frac{|(V_{KM})_{us}|}{|(V_{KM})_{td}|}&\frac{(V_{KM})_{ud}|}{|(V_{KM})_{cs}|}
\end{array}
\right ) J \;,\;\;A' = \frac{1}{2}  \left ( \begin{array}{lll}
\frac{|(V_{KM})_{cs}|}{|(V_{KM})_{tb}|}&\frac{|(V_{KM})_{ub}|}{|(V_{KM})_{ts}|}&\frac{|(V_{KM})_{ub}|}{|(V_{KM})_{cs}|}\\
\frac{|(V_{KM})_{td}|}{|(V_{KM})_{cb}|}&\frac{|(V_{KM})_{ub}|}{|(V_{KM})_{td}|}&\frac{|(V_{KM})_{ub}|}{|(V_{KM})_{cd}|}\\
\frac{|(V_{KM})_{td}|}{|(V_{KM})_{cs}|}&\frac{|(V_{KM})_{td}|}{|(V_{KM})_{us}|}&\frac{|(V_{KM})_{cs}|}{|(V_{KM})_{ud}|}
\end{array}
\right ) J \;,\nonumber\\
&&\\
&&\tilde A = \frac{1}{2}  \left ( \begin{array}{lll}
\frac{|(V_{KM})_{ts}|}{|(V_{KM})_{cb}|}&\frac{|(V_{KM})_{tb}|}{|(V_{KM})_{us}|}&\frac{|(V_{KM})_{us}|}{|(V_{KM})_{cb}|}\\
\frac{|(V_{KM})_{tb}|}{|(V_{KM})_{cd}|}&\frac{|(V_{KM})_{tb}|}{|(V_{KM})_{ud}|}&\frac{|(V_{KM})_{ud}|}{|(V_{KM})_{cb}|}\\
\frac{|(V_{KM})_{cd}|}{|(V_{KM})_{ts}|}&\frac{|(V_{KM})_{ud}|}{|(V_{KM})_{ts}|}&\frac{|(V_{KM})_{us}|}{|(V_{KM})_{cd}|}
\end{array}
\right )J\;,\;\;\tilde A' = \frac{1}{2}  \left ( \begin{array}{lll}
\frac{|(V_{KM})_{cb}|}{|(V_{KM})_{ts}|}&\frac{|(V_{KM})_{us}|}{|(V_{KM})_{tb}|}&\frac{|(V_{KM})_{cb}|}{|(V_{KM})_{us}|}\\
\frac{|(V_{KM})_{cd}|}{|(V_{KM})_{tb}|}&\frac{|(V_{KM})_{ud}|}{|(V_{KM})_{tb}|}&\frac{|(V_{KM})_{cb}|}{|(V_{KM})_{ud}|}\\
\frac{|(V_{KM})_{ts}|}{|(V_{KM})_{cd}|}&\frac{|(V_{KM})_{ts}|}{|(V_{KM})_{ud}|}&\frac{|(V_{KM})_{cd}|}{|(V_{KM})_{us}|}
\end{array}
\right ) J\;.\nonumber
\end{eqnarray}

\bigskip

One immediately finds an invariant quantity: $I_{UB}= A_{ij}A'_{ij} = \tilde A_{ij} \tilde'_{ij} = J^2/4$. Here no summations on $i$ and $j$.
If this quantity is zero, there is no CP violation. The universal nature of the Jarlskog parameter $J$ is also present in the $UB$ representation.

\bigskip

\noindent
{\bf  Unitarity Boomerang and New Physics}

We now discuss how possible new physics information may show up in the $UB$ analysis.
There are different ways the $UBs$ can be used to hunt for new physics.
We will discuss two ways to detect new physics beyond the three generation KM model.

\bigskip

One of them is to see if a $UB$ can be formed as expected after the relevant sides, such as the sides
shown in Fig.1 are measured. The construction of the $UB$ uses the property that there is a common angle. With this constraint, if there is new physics to change the length of the sides in a fashion which is not a universal scaling, CP conserving or violating,  then one cannot
close the $UB$. Another way of presenting this situation is that, if one constructs the $UT(2)$ and $UT(2')$ first, then there may not be a common inner angle.

\bigskip

The above possibility may reveal information which cannot be obtained using only one $UT$. An example in which this might happen is
the simple extension to four generation of quarks with the addition of $t'$ and $b'$. In the case $V_{t'd} = 0$, the defining equation, Eq.(\ref{ut1}),
for $UT(2)$ is still the same and one can define effective inner angles and sides. If one just checks whether the sum of inner angles is $180^\circ$ from direct measurements of individual angles, and the angles determined by the sides, they are consistent. No sign of
new physics will show up by this analysis. However, the $UT(2')$ may be modified by a new term $V_{ub'}V^*_{tb'}$. Only information on $UT(2')$ is also compared with that from $UT(2)$, it is possible to decide if the type of new physics described above exists. The $UB$ analysis contains this comparison together all in one.

\bigskip

Another is to use the property of the invariant quantity $I_{UB}$ discussed earlier. It must equal to $J^2/4$. Taking the square root, one can check
with one of the $UT$ areas in the same $UB$. If more than one $UB$ are constructed, one can also compare if their corresponding invariant quantities are equal.

\bigskip

\noindent
{\bf Discussion}

Although the unitarity triangle carries information, about the Kobayashi-Maskawa quark mixing matrix, it explicitly
contains just three parameters, which is one short to completely fix the KM matrix. The unitarity
boomerangs formed using two $UTs$, with a common inner angle, can completely determine the KM matrix, and therefore
better represents quark mixing information. Out of the six possible $UTs$, there are 9 different ways to have a common
inner angle from two $UTs$. In addition, one can align the two longer sides or one longer and one shorter side to form a $UB$.
Therefore there are total 18 possible $UBs$. By studying the unitarity boomerangs, one can obtain all the information enshrined in KM matrix.
We find that although the $UBs$ have different areas, there is an invariant quantity for
all $UBs$ which is equal to a quarter of the Jarlskog parameter $J$ squared. This is an universal representation of CP violation in the
$UB$ framework. We have also discussed how new physics can be hunted for by using information from $UBs$.

\bigskip

As far as graphic representation of the KM matrix, the proposal, to move from a single triangle to a boomerang combination,
reflects, more than anything else, the impressive precision which has been attained  by high-energy experiment.
A unitarity boomerang contains all information of the KM matrix.

\bigskip

If new physics exists which modifies the unitarity nature, deviations from the expected $UB$ can exist. The $UB$ construction of the mixing matrix elements
can also return, with information about new physics.
\bigskip

This work was supported by the World Premier International Research Center Initiative
(WPI initiative) MEXT, Japan. The work of P.H.F. was also supported by U.S. Department
of Energy Grant No. DE-FG02-05ER41418. The work of X.G.H was supported by the NSC and NCTS of ROC.


\end{document}